\documentclass[twocolumn,fp]{jpsj3}
\usepackage{txfonts}
\usepackage{cite,overcite}
\usepackage{amsmath,amssymb,bm,graphicx
}
\usepackage[dvipdfmx]{graphicx}
\usepackage{latexsym}
\usepackage{braket}
\usepackage{mathrsfs}
\usepackage{cancel}
\usepackage{color}
\usepackage{multirow}
\usepackage{mathtools}
\usepackage{mathrsfs}
\usepackage{comment}
\usepackage{cancel}
\renewcommand{\d}{{\rm d}}
\renewcommand{\i}{{\rm i}}
\newcommand{\e}{{\rm e}}

\renewcommand{\Im}{{\rm Im}\,}

\usepackage{xcolor}
\title{Electronic polarization in non-Bloch band theory}

\author{Shohei Masuda and Masaaki Nakamura}

\inst{Department of Physics, Ehime University Bunkyo-cho 2-5, Matsuyama, Ehime 790-8577}

\abst{
Hermitian topological materials are characterized by the nontrivial
relation between topological numbers and edge modes, i.e. the
bulk-boundary correspondence. In non-Hermitian systems, the conventional
correspondence breaks down. Instead, in the non-Hermitian
Su-Schrieffer-Heeger model, the non-Bloch bulk-boundary correspondence,
which is the relation between the non-Bloch winding number and the
non-Hermitian skin effect, is proposed by S. Yao and Z. Wang. We
introduce the non-Bloch polarization as a topological quantity to detect
the non-Hermitian skin effect. Moreover, we also discuss the non-Bloch
bulk-boundary correspondence in two-dimensional systems using the
non-Bloch polarization with spiral boundary conditions.
}

\begin{document}

\date{\today}

\maketitle

\section{Introduction}
%
%
In condensed matter physics, topological insulators have been
extensively studied from the theoretical and experimental aspects
\cite{Haldane1988,Kane-M,Fu-K-M,Bernevig-Z,Bernevig-H-Z2006,
Schnyder-R-F-L,Ryu-S-F-L,Konig-2007,Konig-2008}.  As the central
concept, there is the bulk-boundary correspondence, which is one-to-one
correspondence between a non-zero topological number (bulk) and an
existence of edge modes (boundary) \cite{Schnyder-R-F-L,Ryu-S-F-L}. In
recent years, topological invariants of non-Hermitian systems, which are
described by the non-Hermitian Hamiltonian, has been studied
\cite{Esaki-S-H-K2011,Lee2016,Shen-Z-F2018,Yao-W2018,Yao-S-W,
Yokomizo-M2019,Kawabata-O-S,Gong-A-K-T-H-U,Kawabata-S-U-S,
Okuma-K-S-S,Borgnia-K-S,Vecsei-D-N-S,Li-Mu-L-G,Kunst-E-B-B2018,
Edvardsson-K-Y-B2020,Lee-L-Y}. However, since open spectra in finite
systems with non-Hermiticity are greatly different from eigenvalues of
the corresponding Bloch Hamiltonian, the conventional bulk-boundary
correspondence is broken. This breakdown is resolved by the non-Bloch
band theory, and its relevance was demonstrated in the non-Hermitian
Su-Schrieffer-Heeger (SSH) model and the two-dimensional Wilson-Dirac
(WD) model with imaginary Zeeman fields
\cite{Yao-W2018,Yao-S-W,Yokomizo-M2019}.  The macroscopic number of
eigenstates in these models are localized on the boundary. This is
called as the non-Hermitian skin effect and considered as a kind of edge
modes.


In this paper, as an index to characterize topological invariants of
non-Hermitian lattice system with skin effect, we consider the
electronic polarization which was introduced by Resta to characterize
insulating states \cite{Resta1994,Resta,Resta-S1999,Resta2000}.  This
quantity is given as the expectation value of the exponential position
operator in periodic systems as
$z^{(q)}=\braket{\Psi_0|\e^{\i(2q\pi/L)\hat{X}}|\Psi_0}$ with $q=1$,
where $\hat{X}=\sum_{j=1}^{L} \hat{x}_j$ with $\hat{x}_j$ being the
position operator at $j$-th site, $L$ is the number of lattices, and
$\ket{\Psi_0}$ is the ground-state many-body wave function
\cite{Resta1994,Resta}.  This can also be interpreted as the overlap
between the ground state and a variational excited state appearing in
the Lieb-Schultz-Mattis (LSM) theorem
\cite{Lieb-S-M,Affleck-L,Affleck}. $z^{(q)}$ with $q\geq 2$ can be
interpreted as extensions of the LSM theorem for general magnetizations
and fillings \cite{Oshikawa-Y-A,Yamanaka-O-A}, and by an equivalent
discussion \cite{Aligia-O}.  Although Resta related $z^{(1)}$ with the
electronic polarization as $\lim_{L\to\infty}(e/2\pi)\Im\ln z^{(1)}$,
hereafter we call $z^{(q)}$ itself ``polarization.''

The polarization $z^{(q)}$ has been calculated in various systems
\cite{Resta-S1999,Aligia-O,Nakamura-V,Nakamura-T} and is shown to
characterize topological phases and topological transitions in
one-dimensional (1D) systems.  Recently, an extension of the
polarization to more than two-dimensional (2D) systems has been proposed
based on spiral boundary conditions (SBCs)
\cite{Nakamura-M-N}. Furthermore, the polarization has also been
extended to non-Hermitian systems with periodic boundary conditions
(PBCs) using the biorthogonal basis \cite{Lee-L-Y,Masuda-N}. However,
this extension is not relevant to describe the non-Hermitian skin
effect, because of the breakdown of the bulk-boundary correspondence in
open boundary systems.  Therefore, to resolve this problem, we discuss
the non-Hermitian skin effect in 1D and 2D systems by applying the
concept of the non-Bloch band theory and SBCs to the electronic
polarization.

\section{Polarization}
We consider the 1D tight-binding model only with short-range
hopping. The Hamiltonian is written as
\begin{equation}
 \mathcal{H}=\sum_{ij}c^{\dag}_{i}H_{ij}c^{\mathstrut}_{j},
\end{equation}
where $c^{\dag}_{i}=(c^{\dag}_{i,\alpha_{1}}, c^{\dag}_{i,\alpha_{2}},
\cdots, c^{\dag}_{i,\alpha_{2N/L}})$, $c^{\dag}_{i,\alpha}$ is a
creation operator of a fermion in an orbital $\alpha$ at the $i$-th unit
cell, $N$ is a number of fermions, and $L$ is a number of unit
cells. Here, we assume that the corresponding Bloch Hamiltonian
satisfies the following eigenvalue equations in the biorthogonal basis:
\begin{equation}
 H(k)\ket{u^{\rm R}_{k\mu}}
  =\varepsilon_{\mu}^{\mathstrut}(k)\ket{u^{\rm R}_{k\mu}},\quad
  H^{\dag}(k)\ket{u^{\rm L}_{k\mu}}
  =\varepsilon_{\mu}^{\ast}(k)\ket{u^{\rm L}_{k\mu}},
\label{biorthogonal_system}
\end{equation}
where $\mu$ is a band index, and the normalization condition is
$\braket{u^{\rm L}_{k\mu}|u^{\rm R}_{k\nu}}=\delta_{\mu\nu}$.

The electronic polarization in non-Hermitian systems with PBCs is
defined as follows \cite{Lee-L-Y}:
\begin{equation}
 z^{(q)}=\braket{\Psi_{\rm L}|U^{q}|\Psi_{\rm R}},\quad
  U=\exp\left[\,\i\frac{2\pi}{L}\sum_{j=1}^{L}j n_{j}\,\right],
  \label{z-def}
\end{equation}
where $\ket{\Psi_{\rm R(L)}}$ is the Slater determinant constructed by
the occupied right (left) eigenstates, and
$n_{j}=c^{\dag}_{j}c^{\mathstrut}_{j}$. According to
Ref.\,\citen{Masuda-N}, this is reduced to the following form in
the Bloch space,
\begin{equation}
 z^{(q)}=(-1)^{qN/L}\prod_{n=0}^{L-1}\,\prod_{\mu \in {\rm occupied}}
  \braket{u^{\rm L}_{k_{n+q}\mu}|u^{\rm R}_{k_{n}\mu}},
  \label{reduced-z}
\end{equation}
where $\ket{u^{\rm R(L)}_{k\mu}}$ are given by
Eq.\,\eqref{biorthogonal_system}, $k_n = (2\pi/L)(n+\theta/2\pi)$ for
twisted boundary conditions $c_{j+L}=\e^{\i \theta}c_{j}$.

\section{1D non-Hermitian skin effect}
\subsection{Hamiltonian}

\begin{figure}[t]
\centering
\includegraphics[width=80mm,pagebox=cropbox,clip]{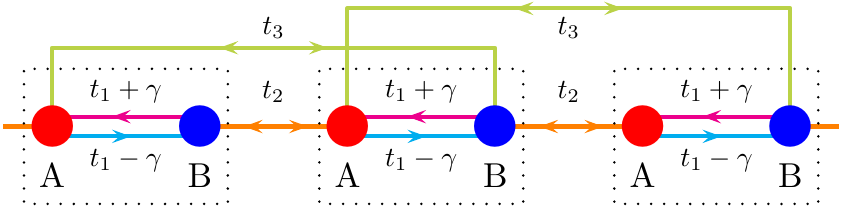}
\caption{
(Color online)
The hopping amplitudes of the non-Hermitian SSH model. The dotted boxes
indicate the unit cells. For $t_3\neq 0$, the eigenvalue equation
becomes quartic form, and the generalized Brillouin zone is not
necessarily a simple circle in the complex $\beta$ plain (see
Fig.\,\ref{fig:gbz_beta_plain}).
} \label{fig:SSH}
\end{figure}

As a model which illustrates the non-Hermitian skin effect, we consider
the non-Hermitian SSH model.  \cite{Yao-W2018,Yokomizo-M2019} The
Hamiltonian of this system is given by (see Fig.\,\ref{fig:SSH})
\begin{align}
 \mathcal{H}&=t_{1} \sum_{j}
 c^\dag_{j,{\rm A}} c^{\mathstrut}_{j,{\rm B}}
 +t_{2} \sum_{j} c^\dag_{j,{\rm B}} c^{\mathstrut}_{j+1,{\rm A}}
 + {\rm H.c.} \notag\\
 &\ + \gamma \sum_{j} c^\dag_{j,{\rm A}} c^{\mathstrut}_{j,{\rm B}}
 - {\rm H.c.}\notag\\
 &\ + t_{3} \sum_{j} c^\dag_{j,{\rm A}} c^{\mathstrut}_{j+1,{\rm B}}
 +{\rm H.c.}.
 \label{SSH_real-space}
\end{align}
The Bloch Hamiltonian is
\begin{equation}
 H(k)=
  (t_{1}+t_{2}\cos k)\tau_{x}+(t_{2}\sin k+\i \gamma)\tau_{y},
  \label{SSH_k-space}
\end{equation}
where $\tau_{\mu}\,(\mu=x,y,z)$ are Pauli matrices. Then the eigenvalue
is given by
\begin{align}
 \varepsilon_{\pm}(k)
 &=\pm\sqrt{t_{1}^2+t_{2}^2-\gamma^2+2t_{2}(t_{1}\cos k+\i\gamma\sin k)}.
\end{align}
The energy gap closes at $t_{1}=\pm(t_{2}\pm \gamma)$.  When $\gamma=0$,
this model is Hermitian, and phase transitions occur at $t_{1}=\pm
t_{2}$. In this case, the orthogonality of eigenstates excludes this
non-Hermitian skin effect. However, for $\gamma\neq 0$, the system may
show the non-Hermitian skin effect due to the asymmetry of the hopping
amplitude and open boundary conditions.

\subsection{Polarization for $t_{3}=0$}

\begin{figure}[t]
\centering
\includegraphics[width=80mm,pagebox=cropbox,clip]{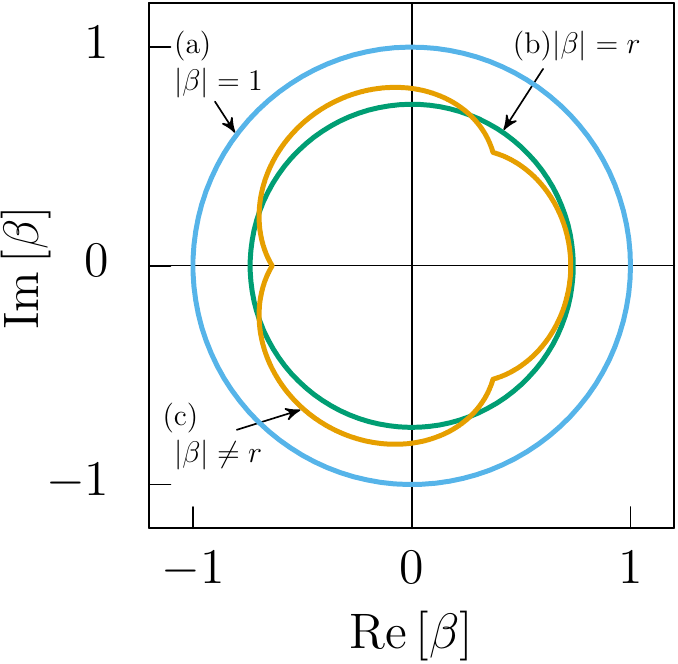}
\caption{
(Color online)
(a) The Brillouin zone for $\gamma=0$ and (b,c) the generalized
Brillouin zone $C_{\beta}$ plotted in the $\beta$ plain.  Parameters are
(b) $t_{1}=1.7, t_{2}=1,t_{3}=0, \gamma=0.5$, and (c)
$t_{1}=0.3,t_{2}=0.8,t_{3}=0.2, \gamma=0.5$.  }
\label{fig:gbz_beta_plain}
\end{figure}

In order to see the non-Hermitian skin effect, we apply the procedure of
the non-Bloch band theory to the Bloch
Hamiltonian \cite{Yao-W2018,Yokomizo-M2019}. By the replacement $\beta=\e^{\i
k}\,(k\in\mathbb{C})$ in Eq.\,\eqref{SSH_k-space}, we obtain the
non-Bloch Hamiltonian as
\begin{equation}
 H(\beta)=
  \begin{pmatrix}
   &h_{+}(\beta) \\
   h_{-}(\beta)&
  \end{pmatrix},
  \label{SSH_beta}
\end{equation}
where $h_{\pm}(\beta)=t_{1}\pm\gamma+t_{2}\beta^{\mp1}$. In the
non-Bloch band theory, we treat the eigenvalue equation ${\rm
det\,}[H(\beta)-E]=0$ as the algebraic equation in terms of
$\beta$. Considering the condition $|\beta_{1}|=|\beta_{2}|$, where
$\beta_{1,2}$ are solutions of the eigenvalue equation, we get the
generalized Brillouin zone $C_{\beta}$:
\begin{equation}
 \beta_{1,2}=r\e^{\i \tilde{k}},
\end{equation}
where $r=\sqrt{|(t_{1}-\gamma)/(t_{1}+\gamma)|}$, and
$\tilde{k}\in\mathbb{R}$.  For $t_3\neq 0$, the eigenvalue equation
becomes quartic form, and the generalized Brillouin zone $C_{\beta}$ is
not necessarily a simple circle in the complex $\beta$ plain (see
Fig.\,\ref{fig:gbz_beta_plain}).

The polarization which is given by Eq.\,\eqref{reduced-z} is extended to
the non-Bloch Hamiltonian using $\tilde{k}$ instead of $k$, which is
discretized as $\tilde{k}=(2\pi/N_{\beta})n$ for
$n=0,1,\cdots,N_{\beta}-1$:
\begin{equation}
 z^{(1)}_{\beta}=(-1)\prod_{n=0}^{N_{\beta}-1}\,
  \braket{u^{\rm L}_{\tilde{k}_{n+1}-}|u^{\rm R}_{\tilde{k}_{n}-}},
 \label{reduced-z_beta}
\end{equation}
where we choose $q=1$ because $N=L$ in the present model. Note that
$N_{\beta}$ is not the system size $L$ but the number of grid points on
$C_{\beta}$. Hereafter, we call $z^{(1)}_{\beta}$ ``non-Bloch
polarization.'' It can be shown that $z^{(1)}_{\beta}$ is real by pseudo
Hermiticity of the system.

On the other hand, it is already known that the non-Hermitian skin
effect is characterized by the non-Bloch winding number
\cite{Yao-W2018}. Since Eq.\,\eqref{SSH_beta} has chiral symmetry
$\tau_{z}H(\beta)\tau_{z}=-H(\beta)$, the non-Bloch winding number is
introduced as
\begin{equation}
 w=\frac{\i}{2\pi}\int_{C_{\beta}}\d s \, s^{-1},
  \label{w-def}
\end{equation}
where 
$s(\beta)=h_{+}(\beta)/\sqrt{h_{+}(\beta)h_{-}(\beta)}$.  As discussed
in Ref.\,\citen{Ryu-S-F-L}, there is a correspondence between the
winding number $w$ and the polarization $z$ in Hermitian systems.
Therefore the same relation is expected to be satisfied for the
non-Bloch polarization as
\begin{equation}
 w=1-\frac{1}{\pi}\Im\ln z^{(1)}_{\beta} 
 \label{w-z_rel}
\end{equation}
for $N_\beta\to\infty$.  It is easy to check this correspondence for
$t_3=0$ where the radius of the generalized Brillouin zone $C_{\beta}$
is a circle with fixed $r$.  However, for $t_3\neq 0$, this relation is
nontrivial.

\begin{figure}[t]
\centering
\includegraphics[width=80mm,pagebox=cropbox,clip]{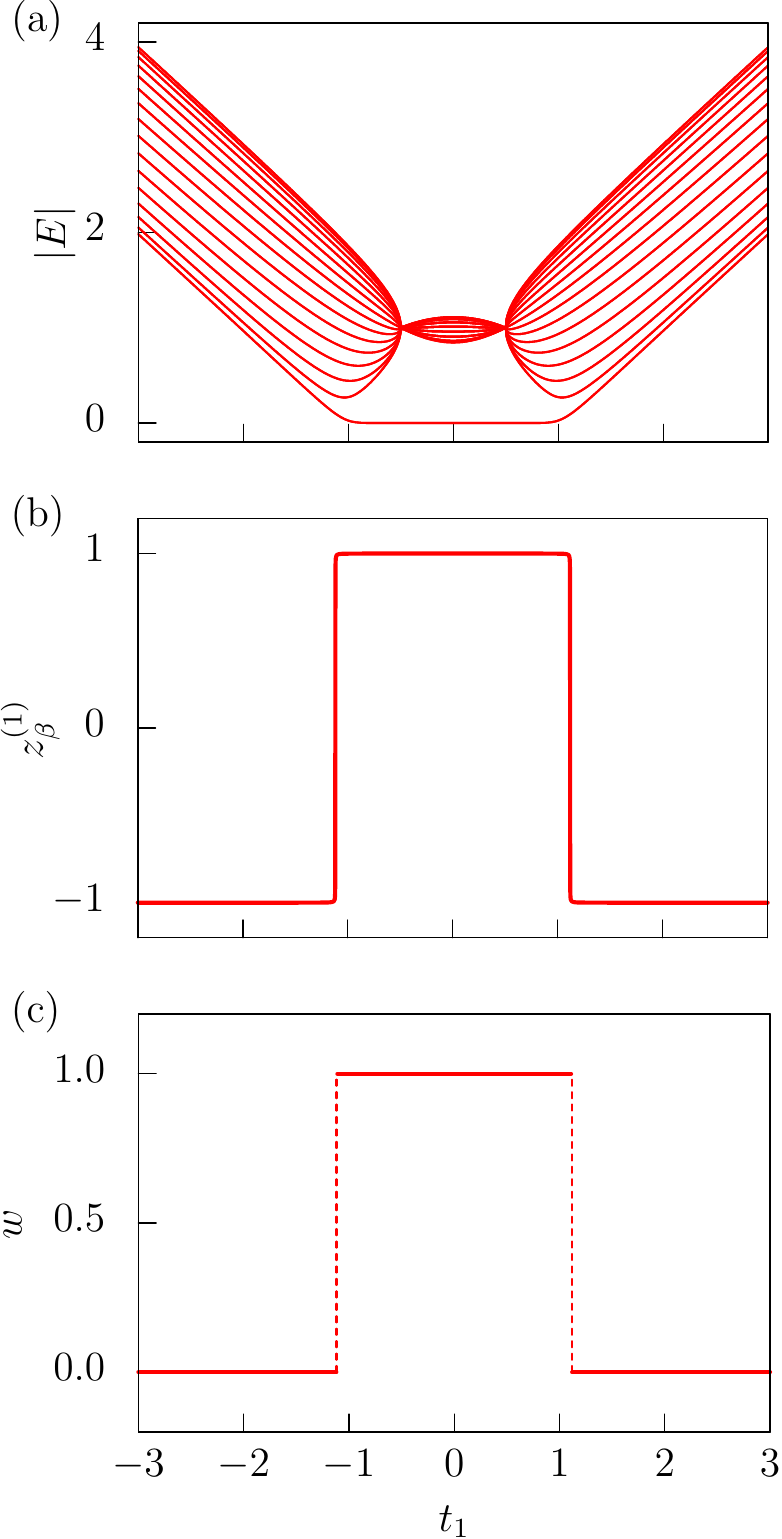}
\caption{
(Color online)
(a) The absolute value of eigenvalues, (b) the non-Bloch polarization
and (c) the non-Bloch winding number in the non-Hermitian SSH model with
$t_{2}=1,\gamma=0.5$ and $t_{3}=0$. The eigenvalues are computed for
$L=16$ with open boundary conditions, and the polarization is calculated
for $N_{\beta}=8192$. Phase transition points are
$|t_{1}|=\sqrt{t_{2}^2+\gamma^2}\simeq1.12$.  For
$|t_1|<\sqrt{t_{2}^2+\gamma^2}$, the zero modes appear in the spectra,
and the winding number takes $w=1$ reflecting the non-Hermitian skin
effect as discussed in Ref.\,\citen{Yao-W2018}. In addition to these,
$z^{(1)}_{\beta}$ changes the sign at the same points.
}\label{fig:gbz_result}
\end{figure}

In Fig.\,\ref{fig:gbz_result}, we show the numerical results for
$t_{3}=0$. We find that the sign of the polarization $z^{(1)}_{\beta}$
defined on the generalized Brillouin zone changes at transition points
$t_{1}=\pm\sqrt{t_{2}^2+\gamma^2}$, and $z^{(1)}_{\beta}=1$ in the
region with the zero modes $|E|=0$ and $w=1$, reflecting the
non-Hermitian skin effect \cite{Yao-W2018}. Whereas $w$ is defined
except at the transition points, $z^{(1)}_{\beta}$ is a continuous
quantity. Therefore, $z^{(1)}_{\beta}$ is more convenient quantity to
detect the phase transition points than $w$, especially in numerical
analysis.

\subsection{Polarization for $t_{3}\neq0$}

For $t_3\neq 0$, we can also calculate the polarization
$z^{(1)}_{\beta}$ and show that the results coincide with those obtained
by the winding number $w$ \cite{Yao-W2018}.  We explain how to find the
generalized Brillouin zone in this case following
Ref.~\citen{Yokomizo-M2019}. We assume that two solutions, which are
$\beta$ and $\beta'=\beta\e^{\i\theta}$, satisfy the eigenvalue
equations:
\begin{equation}
h_{+}(\beta)h_{-}(\beta)=E^2,\quad
h_{+}(\beta\e^{\i \theta})h_{-}(\beta\e^{\i \theta})=E^2,
\end{equation}
where $\theta$ is an arbitrary constant. Then we obtain
\begin{equation}
h_{+}(\beta)h_{-}(\beta)-h_{+}(\beta\e^{\i \theta})h_{-}(\beta\e^{\i \theta})=0,
\end{equation}
which is independent of $E$. By fixing $\theta$, we solve the above
equation and label the four solutions as
$|\beta_{1}|\leq|\beta_{2}|\leq|\beta_{3}|\leq|\beta_{4}|$. According to
the non-Bloch band theory, the trajectory given by the condition
$|\beta_{2}|=|\beta_{3}|$ is the generalized Brillouin zone $C_{\beta}$,
which is not a circle in general (see Fig.\,\ref{fig:gbz_beta_plain}).
%
For $C_{\beta}$ which is a closed loop, we label each $\beta$'s of
$N_{\beta}$ grid points as
\begin{equation}
 0\leq{\rm arg\,}\beta^{(1)}<{\rm arg\,}\beta^{(2)}
  <\cdots<{\rm arg\,}\beta^{(N_{\beta})}<2\pi.
\end{equation}
Then we introduce the polarization defined on $C_{\beta}$ as follows:
\begin{equation}
 z^{(1)}_{\beta}=(-1)\prod_{n=1}^{N_{\beta}}\,
  \braket{u^{\rm L}_{\beta^{(n+1)}-}|u^{\rm R}_{\beta^{(n)}-}},
 \label{reduced-z_gbz}
\end{equation}
where $\ket{u^{\rm R(L)}_{\beta-}}$ is the right (left) eigenstate of
$H(\beta)$, and $\beta^{(N_{\beta}+1)}\equiv \beta^{(1)}$. When
$C_{\beta}$ is a circle, Eq.\,\eqref{reduced-z_gbz} is reduced to
Eq.\,\eqref{reduced-z_beta}.

\begin{figure}[t]
\centering
\includegraphics[width=80mm,pagebox=cropbox,clip]{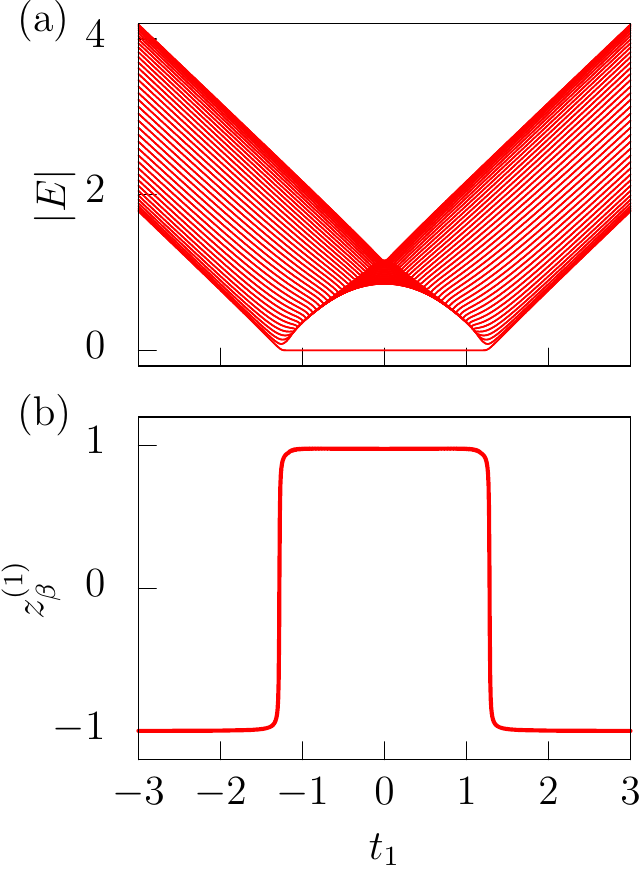}
\caption{
(Color online)
(a) The absolute value of the eigenvalues for $L=40$ and (b)
the polarization for $N_{\beta}=200$ with
$t_{2}=1,t_{3}=0.2,\gamma=0.3$.  $z^{(1)}_{\beta}$ changes the sign at
the same points where the zero modes appear in the spectra. 
} \label{fig:gbz_result_nonzero_t3}
\end{figure}

In Fig.\,\ref{fig:gbz_result_nonzero_t3}, we show the absolute value of
eigenvalues $|E|$ and the polarization $z^{(1)}_{\beta}$ for
$t_{3}\neq0$. In this case, the results also show that the
correspondence exists between the zero modes in the spectra and
$z^{(1)}_{\beta}$. There is also a relation between winding number $w$
and $z^{(1)}_{\beta}$ as Eq.~(\ref{w-z_rel}).

\section{2D non-Hermitian skin effect}
As a model of 2D systems, we consider the WD model \cite{Wilson,Qi-W-Z}
including the non-Hermitian term (imaginary Zeeman field) as follows
\cite{Yao-S-W}:
\begin{align}
 H(\bm{k})=H_{\rm WD}(\bm{k})+\i\sum_{\mu=x,y} \gamma_{\mu}\tau_{\mu},
 \label{nH_WD_lattice}\\
 H_{\rm WD}(\bm{k})=t\sum_{\mu=x,y}\sin k_{\mu}\tau_{\mu}+d_{z}\tau_{z},
\label{WD_lattice}
\end{align}
where
\begin{equation}
d_{z}=M-B\sum_{\mu=x,y}\cos k_{\mu}.
\end{equation}
Then the eigenvalues are given by
\begin{equation}
 \varepsilon_{\pm}(\bm{k})
  =\pm\sqrt{\sum_{\mu=x,y}(t\sin k_{\mu}+\i \gamma_{\mu})^2+d_{z}^2}.
\end{equation}
When $\gamma_{\mu}=0$, this system is Hermitian, and phase transitions
occur at $B=\pm M/2$, where the energy gap closes.

In order to simplify this model, we need the following three steps.
First, we consider the continuum version of Eq.\,\eqref{WD_lattice}
using the Taylor expansion around $k_{\mu}=0$ as,
\begin{align}
 H_{\rm cont.}(\bm{k})
 =&\sum_{\mu=x,y}(t k_{\mu}+\i \gamma_{\mu})\tau_{\mu}\notag\\
 &+\left[M-2B+\frac{B}{2}(k_{x}^2+k_{y}^2)\right]\tau_{z}.
\label{WD_cont}
\end{align}
Next, considering the $\gamma_{\mu}$-term as a perturbation, the
relevant eigenstate of $H_{\rm cont.}$ is exponentially localized which
corresponds to the non-Hermitian skin effect. By the replacement
$k_{\mu}\to \tilde{k}_{\mu}-\i\gamma_{\mu}/t$, we then obtain the
non-Bloch Hamiltonian
\begin{align}
 \tilde{H}_{\rm cont.}(\tilde{\bm{k}})
 =&t\tilde{k}_{x}\tau_{x}+t\tilde{k}_{y}\tau_{y}+\tilde{d}_{z}\tau_{z}\notag\\
 &-\frac{B}{t}\sum_{\mu=x,y}
 \left(\frac{\gamma_{\mu}^2}{2t}+\i \gamma_{\mu}\tilde{k}_{\mu}\right)\tau_{z},
\label{nB_WD_cont}
\end{align}
where
\begin{equation}
 \tilde{d}_{z}=M-B\sum_{\mu=x,y}\left(1-\frac{\tilde{k}_{\mu}^2}{2}\right).
\end{equation}
Finally, we restore the continuum model Eq.\,\eqref{nB_WD_cont} to the
lattice model as follows:
\begin{equation}
 \tilde{H}(\tilde{\bm{k}})=H_{\rm WD}(\tilde{\bm{k}})
  -\frac{B}{t}\sum_{\mu=x,y}
  \left(\frac{\gamma_{\mu}^2}{2t}
   +\i \gamma_{\mu}\sin \tilde{k}_{\mu}\right)\tau_{z},
\label{nB_WD_lattice}
\end{equation}
where $H_{\rm WD}$ is defined by Eq.\,\eqref{WD_lattice}, and we set
$\gamma_x=\gamma_y=\gamma$.  The energy gap closes at
$B=M/(2+\gamma^2/t^2)$. Hereafter, we analyze the model
(\ref{nB_WD_lattice}) instead of Eq.~(\ref{nH_WD_lattice}).

The phases appearing in the 2D WD model $H_{\rm WD}(\bm{k})$ are
topologically distinguished by the Chern number which is defined on the
conventional Brillouin zone. Instead of this, the phases of
$\tilde{H}(\tilde{\bm{k}})$ are characterized by the non-Bloch Chern
number \cite{Yao-S-W}
\begin{equation}
 \nu=\frac{2\pi}{L_{x}L_{y}}\sum_{\tilde{\bm{k}}}
  \left(\frac{\partial A_{y}(\tilde{\bm{k}})}{\partial \tilde{k}_{x}}
   -\frac{\partial A_{x}(\tilde{\bm{k}})}{\partial \tilde{k}_{y}}\right),
\label{Chern_gbz}
\end{equation}
where $\tilde{k}_{\mu}=(2\pi/L_{\mu})n_{\mu}$ for
$(n_{\mu}=0,1,\cdots,L_{\mu}-1)$, and $A_{\mu}(\tilde{\bm{k}})=
-\i\braket{u^{\rm L}_{\tilde{\bm{k}}-}| \partial_{\tilde{k}_{\mu}}u^{\rm
R}_{\tilde{\bm{k}}-}}$.

\begin{figure}[t]
\centering
 \includegraphics[width=80mm,pagebox=cropbox,clip]{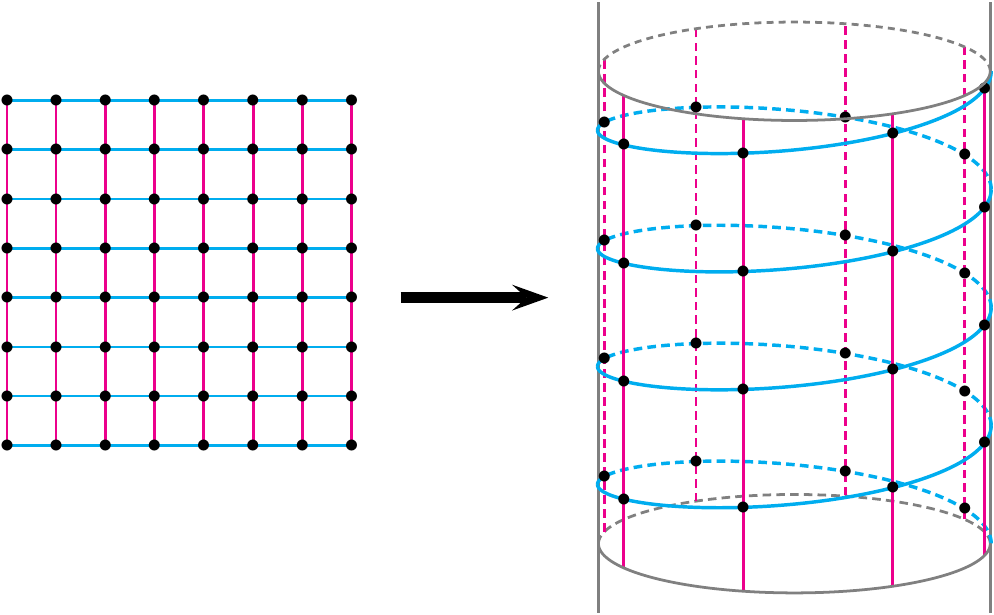}
\caption{
(Color online)
SBCs for a 2D square lattice where the system is represented as a
projected 1D chain (blue lines). The parameter $\Lambda$ is the hopping
range $c_{i+\Lambda,\alpha}^{\dag}c_{i,\alpha}^{\mathstrut}+\mbox{H.c.}$
of the 1D chain originating from the hopping along the $y$ direction
(magenta lines). This is also related to the circumference of the torus.
}\label{fig:SBC}
\end{figure}
\begin{figure}[t]
\centering
 \includegraphics[width=80mm,pagebox=cropbox,clip]{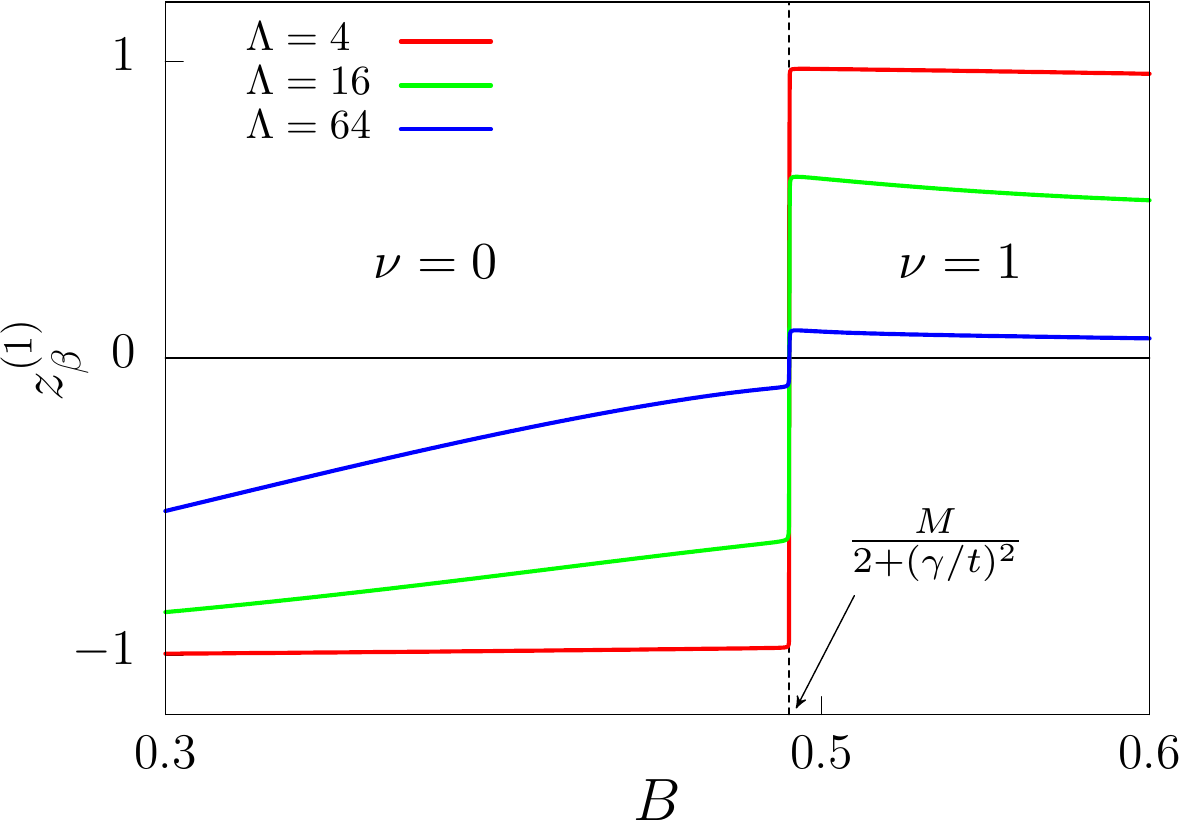}
\caption{
(Color online)
The non-Bloch polarization \eqref{reduced-z_beta} of the effective
lattice model of the non-Hermitian WD model given by
Eq.\,\eqref{nB_WD_lattice}. The parameters are $t=M=1$,
$\gamma_x=\gamma_y=\gamma=0.2$, $N_{\beta}=2^{20}$, and
$\Lambda=4,16,64$. The transition point is
$B=M/(2+\gamma^2/t^2)\simeq0.490$. The non-Bloch Chern number $\nu$ is
defined by Eq.\,\eqref{Chern_gbz}.
}\label{fig:2d_wd_gbz_result}
\end{figure}

In addition to this, we also introduce the non-Bloch polarization for 2D
systems.  For this purpose, we use SBCs that sweep all lattices in 1D
orders as illustrated in Fig.~\ref{fig:SBC}
\cite{Yao-O,Nakamura-M-N}. When we apply SBCs to 2D lattice models, the
2D wave numbers $(k_{x},k_{y})$ are replaced by $(k,\Lambda k)$, where
$k$ is the wave number of the projected 1D chain, and the parameter
$\Lambda$ is the range of the hopping
$c_{i+\Lambda,\alpha}^{\dag}c_{i,\alpha}^{\mathstrut}+\mbox{H.c.}$
related with a hopping to $y$-direction in the original 2D lattice
\cite{Nakamura-M-N}.  It has been shown that the 2D polarization defined
in this way becomes $z^{(1)}=0$ at the topological transition points
\cite{Nakamura-M-N} that are also ensured by the LSM theorem extended to
higher dimensions \cite{Oshikawa,Hastings,Yao-O}.  Then we can introduce
the 2D non-Bloch polarization $z^{(1)}_{\beta}$ as in the same procedure
as in 1D cases using Eq.~(\ref{reduced-z_beta}).

In Fig.\,\ref{fig:2d_wd_gbz_result}, we show the numerical result of the
non-Bloch polarization $z^{(1)}_{\beta}$ of the effective lattice model
of the non-Hermitian WD model given by Eq.\,\eqref{nB_WD_lattice}. Here,
for the $\tilde{\bm{k}}$ space, we have imposed SBCs for the 2D system
and PBCs for the projected 1D chain to use Eq.\,\eqref{reduced-z_beta}.
For large and fixed number of the grid points $N_{\beta}$, the point
where $z^{(1)}_{\beta}=0$ has very small $\Lambda$ dependence, and
$|z^{(1)}_{\beta}|$ tends to converge to 1 for small $\Lambda$.  In the
present case, $\Lambda$ is nothing but the circumference of the torus.
Therefore it is convenient to calculate $z^{(1)}_{\beta}$ near the
thin-torus region, to detect the transition point.
The non-Bloch Chern number $\nu$ defined by Eq.\,\eqref{Chern_gbz} takes
different values on each sides of the transition point
$B=M/(2+\gamma^2/t^2)$, where the sign of $z^{(1)}_{\beta}$ changes.
Thus we can detect the non-Hermitian skin effect in 2D by the non-Bloch
polarization $z^{(1)}_{\beta}$, similarly to the non-Bloch Chern number
$\nu$.

\section{Summary and discussion}
In summary, we have extended Resta's electronic polarization to
characterize insulating states to the non-Hermitian skin effect in 1D
and 2D systems.  The extension has been done by rewriting the
polarization in the Bloch state and applying the non-Bloch band theory
which extends the wave numbers to the complex plain.  This enables
Resta's formalism for the electronic polarization in periodic systems to
detect non-Hermitian skin effect in open boundary systems. Furthermore,
for 2D systems, we have introduced SBCs which sweep all lattice sites in
1D order. Then the non-Bloch electronic polarization defined in 1D
lattice systems has been extended to 2D systems, and it also detects the
transition point for the non-Hermitian skin effect.

We have demonstrated these arguments in the 1D non-Hermitian SSH model
and the 2D non-Hermitian WD model, and found that the regions of the
skin effect identified by the non-Bloch electronic polarization are
consistent with those obtained by the non-Bloch winding number and the
Chern number, respectively.  We also show the consistency of the results
in more general cases of the non-Hermitian SSH model where the
generalized Brillouin zone $C_{\beta}$ is not necessarily a simple
circle.

There are some works that claim the polarization defined in periodic
systems may detect the phase transition relating the non-Hermitian skin
effect \cite{Lee-L-Y,Ortega-Taberner-R-H,Hu-P-F-Z-M-C}.  In these works,
the numerical results of $\tfrac{1}{\pi}\Im\ln z^{(1)}$ of the
non-Hermitian SSH model with PBCs seem to behave similarly to our result
of Fig.~\ref{fig:gbz_result}(c). However, this coincidence is accidental
one due to the finite-size effect: As shown in our previous work
\cite{Masuda-N}, $z^{(1)}$ of the non-Hermitian SSH with PBCs in the
thermodynamic limit takes three values. This means that a finite region
with $z^{(1)}=0$ exists between two gapped phases with $z^{(1)}=\pm
1$. In the region with $z^{(1)}=0$, Resta's polarization
$\tfrac{1}{\pi}\Im\ln z^{(1)}$ does not have physical meaning as in
gapless cases.  On the other hand, if one extracts the sign of $z^{(1)}$
in finite-size systems by abusing the relation $\tfrac{1}{\pi}\Im\ln
z_L^{(1)}$, then it takes only two values 0 or 1 because $z_L^{(1)}$ is
real and finite.  Thus the phase with $z^{(1)}=0$ is overlooked in the
periodic cases. In the end, we need the present extension of the
electronic polarization by the non-Bloch band theory to deal with the
non-Hermitian skin effect.

There is another topological index ``biorthogonal polarization''
proposed in Refs.~\citen{Kunst-E-B-B2018,Edvardsson-K-Y-B2020}.  This
quantity characterizes the number of zero modes
\cite{Edvardsson-K-Y-B2020}, and relationship to the non-Bloch winding
numbers and the non-Bloch polarization is not fully understood.  Since
this quantity is defined in the real space with open boundary
conditions, our non-Bloch polarization is considered to have an
advantage for calculation in larger size systems.


There is another topological index ``biorthogonal polarization''
proposed in Refs.~\citen{Kunst-E-B-B2018,Edvardsson-K-Y-B2020}.  This
quantity characterizes the number of zero modes,
\cite{Edvardsson-K-Y-B2020}, and relationship to the non-Bloch winding
numbers and the non-Bloch polarization is not fully understood. Since
this quantity is defined in the real space with open boundary
conditions, our non-Bloch polarization is considered to have an
advantage for calculation in larger size systems and for the
comprehension of the bulk-boundary correspondence.

\section{Acknowledgment}
The authors thank N.~Hatano, K.~Imura, and M.~Oshikawa for discussions.
M. N. acknowledges the Visiting Researcher's Program of the Institute
for Solid State Physics, the University of Tokyo, and the research
fellow position of the Institute of Industrial Science, the University
of Tokyo.  M.~N. is supported partly by JSPS KAKENHI Grant Number
20K03769.


\end{document}